\documentclass{sf2a-conf2014}
\usepackage[]{natbib}
\usepackage{graphicx}
\usepackage{epstopdf}
\usepackage{hyperref}
\usepackage{amssymb,amsmath}

\def\BibTeX{{\rm B\kern-.05em{\sc i\kern-.025em b}\kern-.08em
    T\kern-.1667em\lower.7ex\hbox{E}\kern-.125emX}}
\bibpunct{(}{)}{;}{a}{}{,}  

\usepackage{MR_AA}

\begin{document}

\TitreGlobal{SF2A 2014}

\title{Magnetohydrodynamics and Solar Physics}
\runningtitle{Magnetohydrodynamics and Solar Physics}

\author{Michel Rieutord$^{1,}$}\address{Universit\'e de Toulouse; UPS-OMP;
IRAP; Toulouse, France}
\address{CNRS; IRAP; 14, avenue Edouard Belin, F-31400 Toulouse, France}

\setcounter{page}{237}

\maketitle

\begin{abstract}
In this short review, I present some of the recent progresses on the
pending questions of solar physics. These questions let us revisit the
solar wind, the solar dynamo problem, the dynamics of the photosphere
and finally have a glimpse at other solar type stars.  Discussing the
use of direct numerical simulations in solar physics, I show that the
full numerical calculation of the flow in a single supergranule would
require more electric power than the luminosity of the sun itself with
present computer technology.

\end{abstract}

\begin{keywords}
solar physics, magnetohydrodynamics
\end{keywords}

\section{Introduction}

Studying the sun is motivated by many reasons. First, we would like to be
able to explain to the street man, what is the sun, what has been its life
until now, what will be its future, why it has permitted the appearance of
life on Earth and whether it is unique or not in the Universe. These many
reasons should be completed by the questions that stimulate astrophysicists
in their quest of a full understanding of this celestial object. Indeed,
the sun is also the closest star and it is a true self-operating physics
laboratory where we can find conditions that cannot be reached on Earth.

Today the sun seems to be well-known: its fundamental parameters have
been determined with some precision, not reached for any other star,
and thanks to helioseismology, namely thanks to a careful
interpretation of the frequencies of the tiny acoustic vibrations of the
sun, we have also been able to check our calculations of its structure. It
turns out that evolutionary models compare nicely to helioseismic
models. Errors on basic thermodynamic quantities like temperature,
density, pressure are around or less than 1\% \cite[][]{gough_etal96}.

Of course the devil is in the details, and details are not missing on
the sun. The first ``big" detail is certainly its magnetic activity. If
$\alpha-\Omega$ dynamo models allow us to retrieve the basic
oscillation of the solar magnetic field, the understanding of
irregularities of the cyle remains a challenge
\cite[][]{R08}. We understand that the cycle is strongly related
to the differential rotation, but this feature of the dynamics of
the sun still escapes a comprehensive view (although some numerical
simulations can reproduce it -- e.g. \citealt{BT02}). But among the
challenges that the sun prompts to us, we should point out the origin
of the supergranulation. This velocity feature has been known for more
than fifty years \cite[e.g.][]{RR10}, and we are still looking for the
reasons of its existence. Last but not least, the problem of heating
the sun's corona is still a pending challenge.

\begin{figure}[t]
\centerline{\includegraphics[width=0.75\linewidth,angle=0]{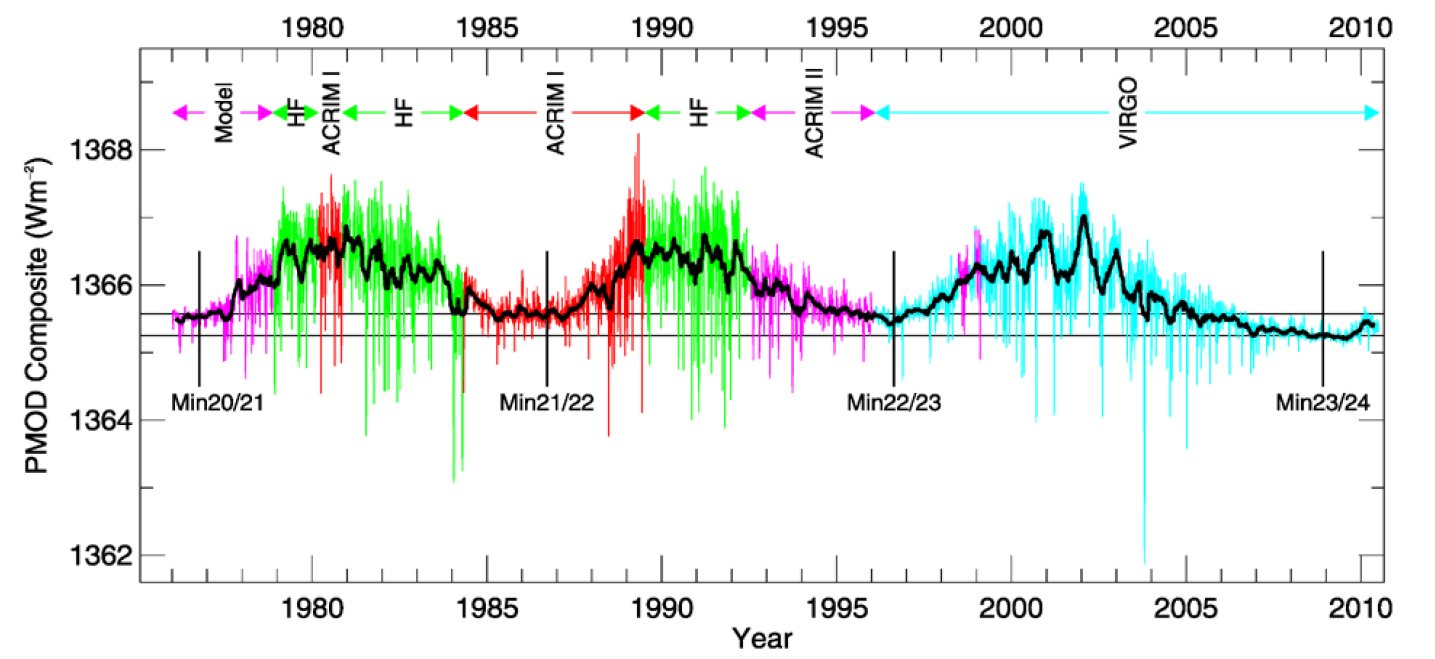}}
\caption[]{Variation of the irradiance with time showing the imprint of
the cycle (credit \citealt{frohlich13}).}
\label{irradiance}
\end{figure}

These questions are actually important for human activities. It is
indeed observed that the magnetic activity of the sun is related to
the irradiance of the Earth (see figure~\ref{irradiance}) and it is
believed that the rather cooler climate that happened in Europe in the period
1645-1715 is actually a consequence of the vanishing solar activity
during that period (the so-called Maunder minimum e.g. \citealt{RNR93}
or \citealt{BTW98}). Of course all the present space activities are
dependent on the particle flux emitted by the sun and should be protected
against the coronal mass ejection. However,  the magnetic field of the
magnetically active sun is also a shield that prevents, in part, the galactic
cosmic rays from reaching the Earth. This is an everyday life concern
for aircraft pilots who face the gamma ray bath due to these cosmic
rays \cite[e.g.][]{obrien_etal96}.

\section{The sun as a laboratory}

The sun is a laboratory where we can observe matter in very extreme
conditions compared to the terrestrial ones. In the old times, this
allowed the discovery of helium by Janssen and Lockyer in 1868. More
recently, the neutrinos oscillations have been discovered in the
neutrinos emitted by the sun \cite[e.g.][]{fukuda_etal98,gough03}. Yet,
the sun is also a laboratory of giant size for studying turbulent
fluid flows, or flows governed by magnetic fields, etc.

In particle physics, we are still looking for the theory that would
unify, for instance, gravitation and the quantum world. In
magnetohydrodynamics, the equations are well-know, namely,

\greq
\displaystyle{\rho \Dt{\vv} = -\na P + \mu\Delta\vv + \vj\wedge\vB}\\
\\
\Div\vv=0 \\
\displaystyle{\dt{\vB} = \rot(\vv\wedge\vB) -\rot(\eta\rot\vB)}\\
\Div\vB=0 \\
\displaystyle{\rho\Dt{e}  =  \Div(\khi\na T) - P\Div\vv + \frac{\mu}{2}(\na :
\vv)^2 + \zeta(\Div\vv)^2 + \eta(\rot\vB)^2/\mu_0}
\egreq
but their general solutions are still a dream.   

\subsection{Solving for the fluid flows}

Solar flows are characterized by very large Reynolds numbers\footnote{We
recall that the Reynolds number of a flow is the ratio $VL/\nu$ where
$V$ is a typical velocity scale of the flow, $L$ is a typical length
scale of the flow and $\nu$ is the kinematic viscosity of the fluid.}
typically above $10^{10}$. Let us consider in more details the challenge
of computing the evolution of a single solar granule from the sole fluid
mechanics equation. With typical size of 1000~km, a typical velocity
of 1~km/s and a typical kinematic viscosity of 10$^{-3}$ m$^2$/s
\cite[e.g.][]{R08}, the Reynolds number is $10^{12}$. The most energetic
scale of the granule is of the size of the granule itself, namely 1000~km,
and the scale at which viscosity smoothes velocity gradients is
$\RE^{-3/4}$ smaller, namely 1~mm. It is therefore clear that numerical
simulations will never reach such a resolution, at least for two
reasons. First, it is useless: we are not interested in such details,
and it is likely that such details are unimportant. Second, it is
energetically impossible with present computer technology as we shall
see now.

To include the smallest vortices, we need a grid mesh about ten times
smaller than the dissipative structure, thus of size equal to 0.1~mm. Kolmogorov
scaling law predicts that velocity amplitude decreases with the one
third power of the scale. Hence, from 1000~km to 1~mm the velocity
fluctuations have been reduced by a factor $10^3$, thus to 1~m/s. Taking
care of these velocities needs a time step of $10^{-4}$s according to
the Courant-Friedrichs-Lewy criterion ($\delta t \leq \delta x/V$). To
summarize, we need a box of size 1000~km with a grid mesh of 0.1~mm, that is
10$^{30}$ grid points. The time step needs to be not larger than
$10^{-4}$~s, so as to follow the flow in real time.

As for the code, we take the PENCIL code as an example \cite[][]{BD02}.
This code needs, typically, 80 floating-point operations
per time step per grid point. Thus, just to follow the sun on one of its
granule, we would need a calculator with that produces $8\times10^{35}$
flops, a number to be compared with the present most powerful machine that
produces $4\times10^{16}$ flops. The difference is enormous, but the
problem is that of the needed energy to run $8\times10^{35}$ flops with
present technology. Such technology indeed can produce 75~Gflops per watt.
Hence, the power needed would be of order $10^{25}$ watts =
0.025~L$_\odot$ for a single granule! A single supergranule that
contains a few hundred granules would need
more than the power of the sun to be computed!
Some colleagues mentioned to me the use of the
quantum computer which may revolutionize the power needed for each
flop, but it is not obvious that every algorithm will benefit from
the efficiency of this computer.

The conclusion of this digression is that the modeling of the subgrid
scales in turbulent flows remains a priority if we wish reasonable
models of solar (and more generally of astrophysical) flows.

\subsection{Three kinds of flows}

As far as we know, turbulence modelling is not universal and therefore
various and documented situations offer useful playgrounds to progress
in our understanding of turbulent flows. As far as the sun is concerned,
three regions may be observed and may lead to new guiding lines for
turbulence modeling.

\begin{figure}[t]
\centerline{\includegraphics[width=0.5\linewidth,angle=0]{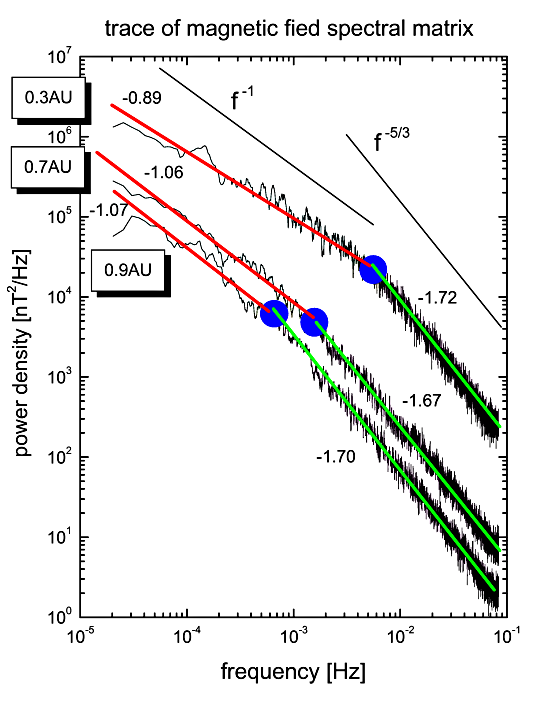}}
\caption[]{Magnetic energy spectra as observed by Helios 2 in 1976
\cite[from][]{BC05}.}
\label{swsp}
\end{figure}

The first one may be the solar wind. This flow has been observed in
situ by many space missions and celebrated spectra of the magnetic field
fluctuations have been measured by Helios 2 (see Fig.~\ref{swsp}). Such
spectra are of interest because they guide us in the difficult problem of
MHD turbulence. For instance, \cite{iroshnikov64} and \cite{kraichnan65}
showed with phenomenological arguments, that the kinetic energy spectrum
should decrease like

\[ E_k\propto k^{-3/2}\]
However this phenomenology was contested by \cite{GS95}
who showed that the anisotropy imposed by the magnetic field is crucial
and therefore that $E_k\propto k_\perp^{-5/3}$, where $k_\perp$ is the wave
vector component orthogonal to the mean-field.

\cite{GM10} and \cite{grappin_etal14} have studied
this problem through turbulence modelling in the spectral space using
an incompressible and perfect (non-diffusive) fluid. The point was to
determine the role of the various parameters that intervene in this
problem. Among other things, they show that the nature of the spectrum,
and therefore its exponent, depends on the intensity of the background
magnetic field and on the correlation time of the large-scale forcing.
They could compute the spectra for various relative angles of the
wave and magnetic vectors, showing the presence of a $k^{-3/2}$
scaling spreading over a decade \cite[][]{grappin_etal14}.

Different conditions may be found at the sun's surface, in the
photosphere. There, the magnetic and velocity fields can both be
measured and spectra obtained \cite[e.g.][]{RR10}, but most detailed
observations are for the velocity fields, thanks to granule tracking
\cite[e.g.][]{RRRD07}. There too, various characteristics of turbulent
flows can be measured. For instance \cite{RMRRBP08} have determined
the first spectrum of surface flows describing supergranulation, while
\cite{RRRMMBZ10} have shown that the supergranulation peak disappears
when a magnetic pore is in the field. In this same study the spectra of
intensity fluctuations have also been derived, showing among other
results that the exponent describing the subgranular scale depends on
the wavelength used for the observation. On the theoretical side, the
main success has certainly been the simulation of the solar photosphere
so as to reproduce the line profiles of various elements and deduce new
constraints on the solar abundances \cite[][]{NSA09}.

\begin{figure}[t]
\centerline{\includegraphics[width=0.75\linewidth,angle=0]{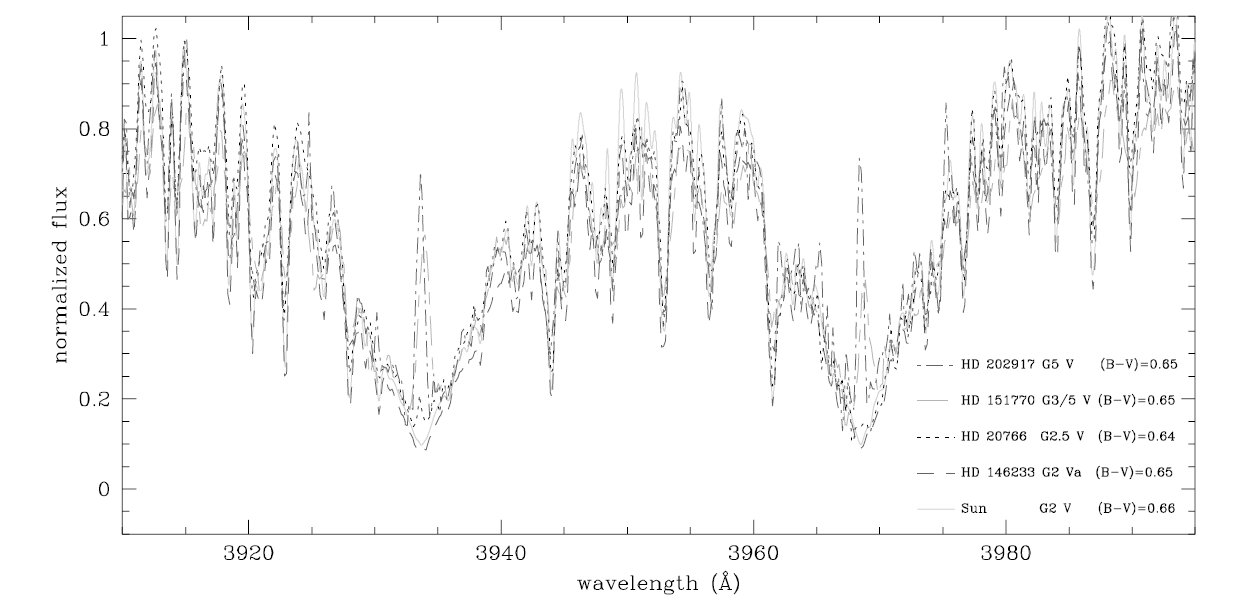}}
\caption[]{The H and K absorption lines of Ca$^+$ for various solar-type
stars \cite[from][]{CM04}. Note the thin emission line that arises in
some stars right in the middle of the large absorption H and K lines.}
\label{HKL}
\end{figure}

In between the solar wind and the solar photosphere are the chromosphere
and the corona. In these regions numerous questions are raised by MHD
phenomena. We cannot avoid mentioning the still pending heating of the
solar corona for which Alfv\'en (or magnetoacoustic) waves are
serious candidates for carrying the energy.
The recent result of \cite{lopezariste_etal13} on the dislocations
observed in propagating MHD waves may both enlight the heating of the
corona and the question of the flux of magnetic helicity at the sun's
surface. Indeed, such dislocations may carry some magnetic helicity and
therefore contribute to the global flux of magnetic helicity at the
surface of the sun. We recall that magnetic helicity, namely

\[ H_m = \intvol \vA\cdot\vB\; dV\]
is an invariant of ideal MHD if the boundary of the fluid does not let
any magnetic flux going through (i.e. if $\vB\cdot\vn=0$ on the boundary).
This is typically an (approximate) invariant of coronal loops. But this
is a quantity that is important to measure so as to estimate its flux at
the surface of the sun. Indeed, one of the recent results of numerical
simulations of fluid dynamos is that saturation of the $\alpha$-effect,
the so-called $\alpha$-quenching is affected by the magnetic helicity.
If magnetic helicity cannot be expelled from the fluid domain, numerical
simulations have shown that the $\alpha$-effect is catastrophically quenched.
At such a low level, this mechanism is no longer effective and would
compromise the solar dynamo \cite[][]{BS05,R08}. The sun manages to
expell this helicity but the process is not well-known.

Hence, measuring the solar flux of magnetic helicity is crucial to put
constraints on the solar dynamo. This is a difficult task that has been
attempted by \cite{dalmasse_etal14} for instance.

\begin{figure}[t]
\centerline{\includegraphics[width=0.7\linewidth,angle=0]{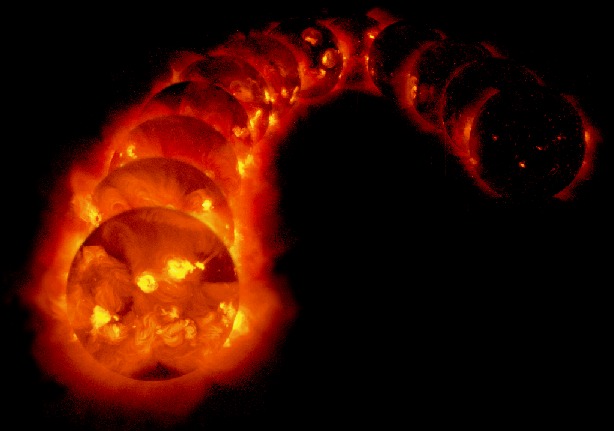}}
\caption[]{The solar magnetic cycle as viewed in the X-rays by the
satellite Yohkoh between 1991 and 2001.}
\label{yokho}
\end{figure}

\section{Moving to other stars}

The sun itself shows only one example of a magnetically active star but
astrophysicists would like a more general picture to appreciate, for
instance, the effects of changing global parameters (mass, age,
rotation, chemical composition, ...) on the magnetic activity.

A longstanding way of monitoring the magnetic activity of stars has
been to measure the intensity of chromospheric lines, especially the H
(396.85 nm) and K (393.368 nm) lines of the calcium ion Ca$^+$ (see
figure~\ref{HKL}).  Understanding the magnetic activity of solar-like
stars has become a crucial point for the detection of exo-planets
because the
magnetic activity raises the detection level of the radial velocity
signature of a planet.  As shown by \cite{livingston_etal07}, the
emission inside the H\&K line is quite nicely correlated to the solar
cycle thus supporting the relevance of this index for monitoring the
activity. Presumably, the emission line inside the large H \& K absorption
lines are coming from the chromosphere of the star but the process of
this emission is not completely clear \cite[][]{hall08}.

Additional difficulties come from the modeling of the corona which is a
region dominated by the magnetic fields. Global models of a
corona like that of the sun are slowly emerging \cite[][]{amari_etal13}.
These model are all the more welcome that the corona is the seat of the
X-ray luminosity of solar-type stars. Such an emission is naturally
another signature of the magnetic activity of stars. In X-rays the sun's
luminosity is quite low, typically,

\[ 10^{-7} L_\odot \leq L_X \leq 10^{-6} L_\odot \]
but is varies with the cycle as nicely shown by the celebrated pictures
obtained with the Yohkoh satellite (see figure~\ref{yokho}).  Since this
X-ray emission is triggered by shock waves driven by flares in the
corona of the stars, simulation of unstable magnetic configurations are
an appropriate tool to investigate the energy released by the associated
flows \cite[][]{pinto_etal14}.

\section{Conclusion}

Back to the sun we may conclude that our star is indeed a gigantic
laboratory for MHD. There is a huge quantity of available data, but it
is quite scattered \cite[][]{R12}. From these data, constraints on various
high Reynolds number flows may be derived. This is a detailed view of
an active low mass star which should lead to understanding how such an
activity influences the star's environment and further constraints the
habitability problem.

\bibliographystyle{aa} 
\bibliography{/home/virgo/tex/biblio/bibnew}

\end{document}